\documentclass[9pt,twocolumn,twoside]{osajnl}
\journal{ol} % Choose journal (ao,jocn,josaa,josab,ol,optica,pr)

\setboolean{shortarticle}{false}
\usepackage{xcolor}

\title{Sub-Doppler spectroscopy of the near-UV Cs atom 6S$_{1/2}$ - 7P$_{1/2}$ transition in a microfabricated vapor cell}

\author[1,*]{Emmanuel Klinger}
\author[1]{Andrei Mursa}
\author[1]{Carlos M. Rivera-Aguilar}
\author[1]{Rémy Vicarini}
\author[1]{Nicolas Passilly}
\author[1]{Rodolphe Boudot}

\affil[1]{FEMTO-ST, CNRS, Universit\'e de Franche-Comt\'e, ENSMM, 26 chemin de l'Epitaphe 25030 Besan\c{c}on Cedex, France}

\affil[*]{Corresponding author: emmanuel.klinger@femto-st.fr}

\begin{abstract}
We report on the characterization of sub-Doppler resonances detected by probing the 6S$_{1/2}$ - 7P$_{1/2}$ transition of Cs atom at 459 nm in a microfabricated vapor cell. The dependence of the sub-Doppler resonance (linewidth, amplitude) on some key experimental parameters, including the laser intensity and the cell temperature, is investigated. These narrow atomic resonances are of interest for high-resolution spectroscopy, instrumentation, and may constitute the basis of a near-UV microcell optical standard.
\end{abstract}

\setboolean{displaycopyright}{true}

\begin{document}
\twocolumn
\maketitle

%\section{Introduction}
Microfabricated (MEMS) alkali vapor cells are at the core of high-precision integrated atomic quantum sensors and devices \cite{Kitching:APR:2018}, such as microwave \cite{Yanagimachi:APL:2020, Martinez:NC:2023, Carle:OE:2023} and optical \cite{Hummon:O:2018,Maurice:OE:2020} clocks, or magnetometers \cite{Shah:Nature:2007, Griffith:OE:2010, Boto:Nature:2018}. They also constitute a key element for the demonstration of chip-scale laser-cooling platforms \cite{McGilligan:APL:2020}, atomic diffractive optical elements \cite{Stern:NC:2019}, optical isolators \cite{Levy:2020}, voltage references \cite{Teale:AVS:2022}, or quantum memories \cite{Treutlein:2023}.\\
The first chip-scale atomic device was a microwave atomic clock \cite{Knappe:APL:2004}. This clock, based on coherent population trapping \cite{Arimondo:1996}, has offered in its industrial and commercialized version an ultra-low size-power-instability budget, impacting a plethora of industrial and scientific applications. Nevertheless, the short-term stability of these clocks is usually limited  at about 10$^{-10}$ at 1\,s by the frequency noise of their vertical-cavity surface-emitting laser while their long-term stability is degraded by light-shift effects \cite{MAH:APL:2022} or buffer-gas induced collisional shifts \cite{Abdullah:APL:2015, Carle:JAP:2023}.\\
Hot vapor MEMS-based optical frequency standards constitute a new generation of miniaturized clocks, with enhanced stability. These references, which keep the benefit of using wafer-scalable and mass-producible vapor cells while preventing ultra-high vacuum technologies and laser cooling, rely on the interaction of hot atoms with two counter-propagating laser beams of same frequency. This configuration provides the detection of sub-Doppler resonances of linewidth ultimately limited by the natural linewidth of the probed transition.\\
%detection of sub-Doppler resonances,
% achieved 
%by
% making interact 
%having a hot atomic vapor
% ensemble
%to interact .\\
Probing the two-photon transition of Rb atom at 778\,nm in a MEMS cell has led to the demonstration of an optical frequency standard with fractional frequency stability of 1.8~$\times$~10$^{-13}$ at 1\,s and approaching the 10$^{-14}$ level after 1000\,s \cite{Newman:OL:2021}. Besides, stability results at the level of 3~$\times$~10$^{-13}$ at 1\,s and below 5\,$\times$\,10$^{-14}$ at 100\,s were achieved with dual-frequency sub-Doppler spectroscopy in a micromachined Cs cell \cite{Gusching:OL:2023}. Hybrid photonic-atomic references that couple MEMS cells and compact optical cavities have been also demonstrated \cite{Zhang:LPR:2020}.\\
The above-mentioned studies exploit optical transitions of alkali atoms in the near-infrared wavelength domain. 
%With
Given the recent advances of low-noise near-ultraviolet (UV) chip-size lasers \cite{Kippenberg:APLP:2022}, 
%the exploration of
investigating atomic transitions in microfabricated cells in this wavelength domain 
%constitutes
represents a stimulating research 
%axis
direction. In Ref.\,\cite{Pate:OL:2023}, sub-Doppler spectroscopy of the $^1\text{S}_0-^1$P$_1$ transition of strontium at 461 nm was reported in a MEMS cell. With this alkaline-earth atom, a challenge is to operate the cell at a high temperature ($\sim$ 300$^{\circ}$C), required to obtain a sufficiently high vapor pressure, while ensuring a long enough cell lifetime. In Ref. \cite{Cs459}, a compact optical standard based on the $6\text{S}_{1/2}-7\text{P}_{1/2}$ transition of Cs atom was reported, achieving a stability of 2.1~$\times$~10$^{-13}$ at 1~s and averaging down to a few 10$^{-14}$. However, this reference was based on a 5-cm long glass-blown cell, not compliant with the advent of a fully-miniaturized and low-power optical clock.\\
In this paper, we report on the characterization of sub-Doppler resonances detected in a microfabricated cell by probing, in a simple saturated absorption configuration, the Cs atom 6S$_{1/2}$~--~7P$_{1/2}$ transition at 459 nm. The impact of the laser intensity and cell temperature on the sub-Doppler resonance is experimentally investigated. Optimal laser intensity and cell temperature values for the development of a near-UV microcell-stabilized frequency reference are identified. With detection-noise measurements we predict a short-term stability in the 10$^{-13}$ range at 1\,s, with a photon shot-noise limit in the low 10$^{-14}$ range.\\
\begin{figure}[t]
\centering
\includegraphics[width=0.96\linewidth]{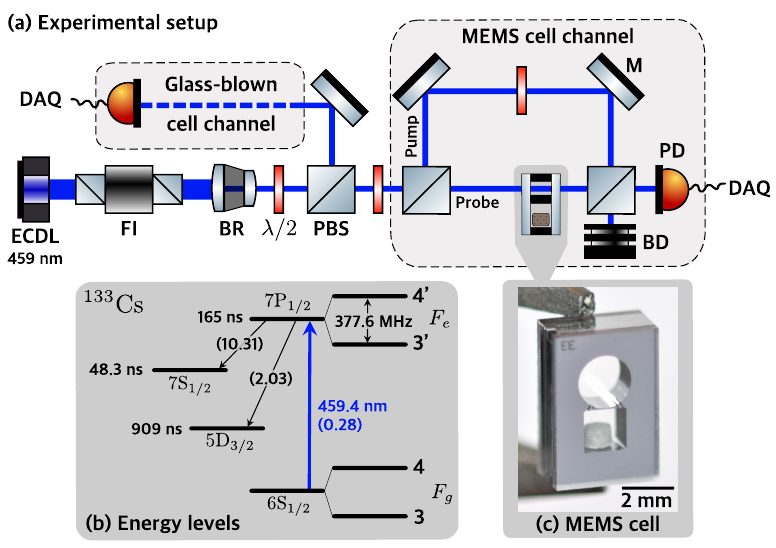}
\caption{(a) Set-up for sub-Doppler spectroscopy of the Cs atom $6S_{1/2} \rightarrow 7P_{1/2}$ transition at 459~nm. ECDL: external-cavity diode laser, FI: Faraday optical isolator, BR: beam reducer, PBS: polarizing beam splitter, M: mirror, $\lambda/2$: half-wave plate, BD: Beam dump, PD: photodiode, DAQ: acquisition card. %For simplicity, 
The glass-blown cell channel
%(identical to that of the MEMS cell),
%is not 
%completely shown
(not shown) is identical to that of the MEMS cell. (b) Energy levels and decay paths involved in the 6S$_{1/2} - 7$P$_{1/2}$ transition
%with parameters 
whose features are extracted from Refs.\,\cite{Safranova:PRA:2016,Williams:LPL:2018,Toh:PRA:2019}. The numbers in parenthesis are the dipole moments given in units of electronic charge times Bohr radius $ea_0$ \cite{Roberts:PRA:2023}. (c) Photograph of the Cs MEMS cell.}
\label{figure1}
\end{figure}
Figure \ref{figure1}(a) shows the experimental setup used to perform sub-Doppler spectroscopy of the $6\text{S}_{1/2} \rightarrow 7\text{P}_{1/2}$ transition in $^{133}$Cs vapor. The laser source is an external-cavity diode laser (ECDL) tuned at 459~nm. An optical isolation stage of 45 dB is placed at the output of the laser head to reduce optical feedback. The intensity of the light is then split such that sub-Doppler spectroscopy, performed in a pump-probe configuration, can be performed in two different cells. The first cell is a
% microfabricated (
MEMS Cs vapor cell \cite{Vicarini:SA:2018}. In this cell, shown in Fig.\,\ref{figure1}(c), atom-light interaction is produced in a 2-mm diameter and 1.4-mm long cylindrical cavity. The cell is placed into a temperature-controlled physics package, surrounded by a magnetic shield. The second cell is an evacuated glass-blown (GB) reference cell with a diameter of 25\,mm and a length of 50\,mm. The latter is also temperature-controlled but not covered by any magnetic shield. For the two cells, the pump and probe powers (intensities) are noted $P_p$ ($I_p$) and $P_b$ ($I_b$), respectively. A set of lenses is used at the laser output 
%such that
to adjust the beam ($1/e^2$) diameter 
%is
to about 2.2\,mm.\\
We started by recording the transmission profiles in the absence of the pump beam for several values of the probe power. The inset in Fig.\,\ref{figure2} shows the transmission spectrum of the $F_g=4\rightarrow 3',4'$ group of transitions recorded with the MEMS cell at a temperature $T_{\text{MEMS}}=117\,^{\circ}$C, and a probe power $P_{b}$ of $5\,\mu$W. As the natural linewidth of the $6\text{S}_{1/2}\rightarrow 7\text{P}_{1/2}$ transition $\Gamma_N/2\pi \sim 0.96\,$MHz \cite{Toh:PRA:2019} is much smaller than the Doppler-width ($\Gamma_D/2\pi\sim 0.7\,$GHz at $25\,^\circ$C), we approximate the transmission profile by a sum of two Gaussian lines, one for each hyperfine transition. We extract here a full-width at half-maximum (FWHM) of about 775\,MHz, and 83\,\% transmission, in agreement with a model of linear absorption \cite{Pizzey:NJP:2022}. This measurement was repeated for several values of the probe power, for the two cells. Results are shown in Fig.\,\ref{figure2}. 
\begin{figure}[htb]
\centering
\includegraphics[scale=0.95]{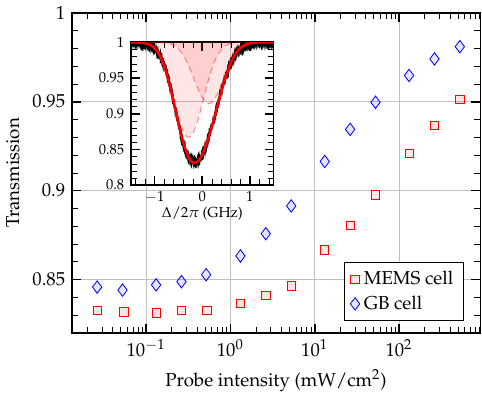}
\caption{Transmission of the cesium $6\text{S}_{1/2}\rightarrow~7\text{P}_{1/2}$~($F_g=4\rightarrow~3',4'$) line center versus probe intensity, for the MEMS and GB cells. The inset shows an example of two Gaussian fits (red) to the data (black), recorded in the MEMS cell. Temperatures of both cells were adjusted to have about the same optical depth: $T_{\text{MEMS}} = 117\,^\circ$C, $T_{\text{GB}} = 61\,^\circ$C.}
\label{figure2}
\end{figure}
\begin{figure}[h!]
\centering
\includegraphics[width = \linewidth]{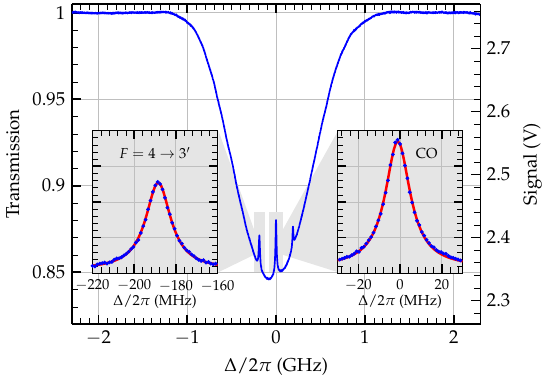}
\caption{Sub-Doppler spectroscopy of the Cs atom 6S$_{1/2} \rightarrow 7$P$_{1/2}$ transition in the Cs MEMS cell. Experimental parameters are $P_{b}= 20\,\mu$W, $P_p=15$\,mW and $T_{\text{MEMS}} = 117\,^{\circ}$C. The two insets show the data (blue dots) and fit (red line) around the $F=4\rightarrow3'$ (left) and crossover (right) resonances.}
\label{figure3}
\end{figure}
Above a probe intensity $I_b$ of
% 50\,\mu$W (
about $1\,$mW/cm$^2$, the transmission starts to increase. This
%is an indication of 
indicates probe-induced saturation of the transitions. We note that the saturation is more pronounced for the GB cell, suggesting that the excited state relaxation is higher in the MEMS cell. In the following, we 
%chose to work at
set $I_b=0.5\,$mW/cm$^{2}$, 
%resulting from
chosen as a trade-off between signal-to-noise ratio and minimal probe broadening.\\
For large enough intensities, the pump burns a hole in the distribution of atomic velocities \cite{Smith:AMJ:2004}. This hole is directly imaged by the (counter-propagating) probe beam, leading to the formation of sub-Doppler peaks with a Lorentzian shape
\begin{equation}\label{eq:amp-subDoppler}
   L(\Delta) = \frac{A}{1+(2\,\Delta/\Gamma)^2}\,,
\end{equation}
where $A$ and $\Gamma$ are the amplitude and the linewidth (FWHM) of the resonance, respectively, while $\Delta= \omega_l-\omega$ is the laser detuning with respect to the angular frequency of the resonance. Figure\,\ref{figure3} shows, in the MEMS cell, sub-Doppler resonances in the bottom of Doppler-broadened profiles. The spectrum was obtained at $T_{\text{MEMS}}=117\,^{\circ}$C with orthogonally-polarized counter-propagating pump and probe beams ($P_p=13\,$mW, $P_{b}=20\,\mu$W). The insets show fits of \eqref{eq:amp-subDoppler} to experimental data for the $F~=~4\rightarrow3'$ and crossover (CO) resonances. The frequency separation of $377.6(2)\,$MHz between the $F=4\rightarrow3'$ and $F=4\rightarrow4'$ transitions \cite{Williams:LPL:2018} was used to calibrate the frequency axis.\\
\begin{figure}[t]
\centering
\includegraphics[scale=0.95]{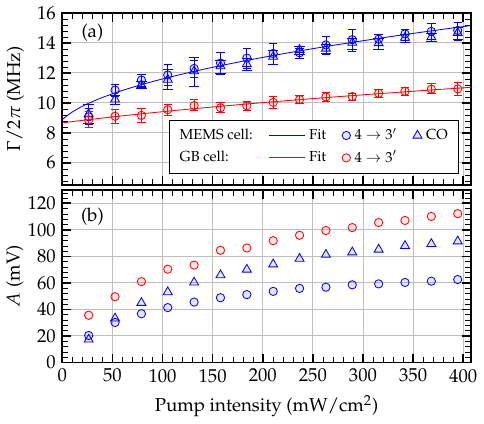}
\caption{Linewidth (a) and amplitude (b) 
of sub-Doppler resonances versus pump light intensity. The error bars correspond to the $2\sigma$ confidence intervals returned by the fit. The probe power is $P_{b} = 20\,\mu$W. The temperature of the MEMS and glass-blown cells are 117$\,^{\circ}$C and  $61\,^{\circ}$C, respectively. Note that the amplitude of the $F=4\rightarrow 4'$ resonance (not shown) was about $1.5\times$ smaller than $4\rightarrow3'$ in both cells.}
\label{figure4}
\end{figure}
The pump intensity leads to power broadening of the sub-Doppler resonances, such that
\begin{equation}\label{eq:linewidthBroad}
    \Gamma' = \Gamma_N\sqrt{1+I_p/I_{sat}} + \Gamma_R\,,
\end{equation}
where $\Gamma_N= 1/2\pi\tau\approx 963\,$kHz is the natural linewidth of the $6\text{S}_{1/2}\rightarrow 7\text{P}_{1/2}$ transition of $^{133}$Cs atoms \cite{Toh:PRA:2019}, and $I_{sat}~=~\epsilon_0 c\hbar^2 \Gamma^2/4|\langle e|d|g\rangle|^2$ is the saturation intensity of a given $F_g\rightarrow F_e$ transition. From the dipole moment value reported in Ref.\,\cite{Roberts:PRA:2023}, we estimate $I_{sat}(6\text{S}\rightarrow7\text{P}_{1/2})\approx 4.9\,$mW/cm$^2$.  The term $\Gamma_R$ comprises all sources of additional broadening (e.g. residual Doppler, collisions, etc.). 
Fitting \eqref{eq:linewidthBroad} to the measured broadening of the sub-Doppler resonances [Fig.\,\ref{figure4}(a)] yields $\Gamma_R~=~7.9(7)\,$MHz.
% , and $I_{sat}=7.3(8)\,$mW/cm$^2$ for the $F=4\rightarrow 3'$ transition, in line with our estimated value of 4.9\,mW/cm$^2$ based on the dipole moment reported in  Ref.\,\cite{Roberts:PRA:2023}.
On the GB cell, we find $\Gamma_R=7.5(6)\,$MHz, consistent with the value observed in Ref. \cite{Cs459} with modulation transfer spectroscopy. With these results, we estimate that at least $0.4\pm1.3\,$MHz are due to collisionnal broadening in the MEMS cell. The difference of slope between the two curves might be explained by differences in sub-level remixing
% process of different origin 
\cite{Steck}. While providing insights into the experimental results, our simple two-level atom model is limited and a complete model accounting for all levels (see e.g. Fig.\,\ref{figure1}) and hyperfine pumping \cite{Smith:AMJ:2004} effects would be necessary to get a complete quantitative understanding. Figure\,\ref{figure4}\,(b) shows that the amplitude of the sub-Doppler resonance is increased with the pump laser intensity. Derived from Figs.\,\ref{figure4}\,(a) and \ref{figure4}\,(b), we find that the amplitude/linewidth ratio of the sub-Doppler resonance
% $2\pi A/\Gamma$
is maximized at the level of about 4.2 mV/MHz for $I_p\approx 320\,$mW/cm$^2$, for the $F=4\rightarrow3'$ resonance. For the crossover resonance, the latter is optimized at the level of 6.3\,mV/MHz for $I_p\approx$~478\,mW/cm$^2$.\\
The cell temperature is also an important parameter to optimize for clocks applications. Figure\,\ref{figure5} shows the amplitude of the sub-Doppler resonance as a function of the MEMS cell temperature, recorded at constant pump and probe powers. The amplitude of the resonance is increased before reaching a plateau at about 122\,$^\circ$C. Although not shown here, note that we did not observe any relevant variation of the resonance linewidth in the tested temperature range, suggesting that the system is not dominated by self-broadening. We found an optimum slope of the sub-Doppler resonance of 4.8\,mV/MHz (8.0\,mV/MHz) around $122\,^\circ$C for the $F=4\rightarrow3'$ (crossover) resonance, respectively. We also note that no degradation of the MEMS cell was observed after six months of operation at temperatures higher than 100\,$^\circ$C.\\
\begin{figure}[t]
\centering
\includegraphics[scale = 1]{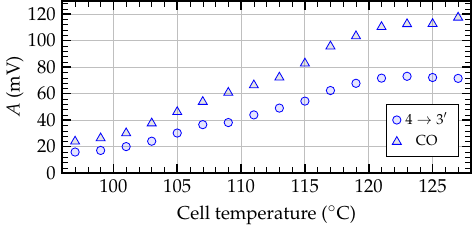}
\caption{Amplitude of the sub-Doppler resonance as a function of the cell temperature, recorded for $P_{b} = 20\,\mu$W, and $P_{p}~=~15\,$mW in the MEMS cell. The circles (triangles) correspond to the $F=3\rightarrow4'$ (CO) respectively.}
\label{figure5}
\end{figure}
%In a last part, we have performed detection noise measurements for predicting what the short-term stability of the microcell-stabilized might be. 
The short-term fractional frequency stability at 1~s of a passive atomic frequency reference is given by
\begin{equation}
   \sigma (1\,\text{s}) \simeq \frac{1}{Q}\cdot \frac{1}{\text{SNR}}\,,
\end{equation} 
where $Q = \nu_0 / \Delta \nu$ is the resonance quality-factor, with $\nu_0$ the laser frequency (6.53\,$\times$\,10$^{14}$\,Hz), $\Delta \nu$ the resonance linewidth, and $\text{SNR = S/N}$ the signal-to-noise ratio in a 1~Hz bandwidth, measured at the modulation frequency $f_M$. Figure\,\ref{figure6} shows the total detection noise measured at half-height of the crossover resonance detected in the MEMS cell at the direct output of the photodiode (Thorlabs PDA36A-EC, with 60 dB gain), with $T$~=~122\,$^{\circ}$C, $P_b$~=~20 $\mu$W and $P_p$~=~18\,mW ($I_p\approx$~470\, mW/cm$^2$). Assuming an operating modulation frequency $f_M$~=~100~kHz, we extract a noise level of $-114$\,dBV$^2$/Hz, i.e. N\,$\simeq$\,2\,$\mu$V. For this measurement, we extract $\Delta \nu$~$\simeq$~16~MHz, $Q$~$\simeq$~4.1~$\times$~10$^7$, $S$~$\simeq$~140\,mV, SNR~$\simeq$~70~$\times$~10$^3$, yielding, in tested conditions, a predicted stability of $\sigma(1~\text{s})\simeq3.5\times 10^{-13}$. As seen in Fig.\,\ref{figure6}, the total detection noise at $f=§100$\,kHz is here limited by the photodiode noise (in the dark). The contribution at 1~s of the photon shot-noise limit \cite{Gusching:JOSAB:2021} 
\begin{equation}
  \sigma_{sn} (1~\text{s}) = \sqrt{\left(\frac{1}{Q}\right)^2 \frac{2 h \nu_0}{C^2 P_o}}\,,
\end{equation}
with $h$ the Planck constant, $P_o~\simeq$ 15~$\mu$W the laser power impinging the photodiode, and $C~\simeq$~0.06 the transmission contrast of the resonance is estimated, in current condition, at 2.4~$\times$~10$^{-14}$.\\
\begin{figure}[t]
\centering
\includegraphics[scale = 1]{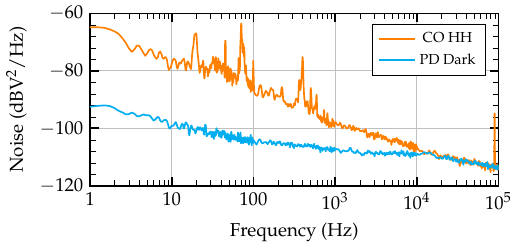}
\caption{Detection noise at the photodiode output for $P_{b}~=~20\,\mu$W, $P_{p}=18\,$mW, and $T_{\text{MEMS}}=122\,^\circ$C, in the MEMS cell, at half-height of the CO resonance. The cyan curve corresponds to the noise of the photodiode in the dark.}
\label{figure6}
\end{figure}
In conclusion, we have reported pump-probe sub-Doppler spectroscopy of the $^{133}$Cs atom $6\text{S}_{1/2}\rightarrow 7\text{P}_{1/2}$ transition in a microfabricated vapor cell. Optimal values of the laser pump intensity and cell temperature have been identified for the demonstration of a near-UV microcell-stabilized laser with a predicted stability at 1\,s in the low-10$^{-13}$ range, %in a 
despite a simple architecture. In the future, we will aim to measure the frequency stability of such a microcell-laser by creating a beatnote between two quasi-identical systems. Microfabricated cells with embedded non-evaporable getters \cite{Boudot:SR:2020} might be also used for reducing the residual pressure of contaminants in the cell and detecting narrower resonances.  

\section*{Funding}
This work has been partly funded by Agence Nationale de la Recherche (ANR) in the frame of the LabeX FIRST-TF (Grant ANR 10-LABX-0048), EquipeX Oscillator-IMP (Grant ANR 11-EQPX-0033) and EIPHI Graduate school (Grant ANR-17-EURE-0002), by R\'egion Bourgogne Franche-Comté, and by Centre National d'Etudes Spatiales (CNES) in the frame of the OSCAR project (Grant 200837/00). The PhD thesis of C. M. Rivera-Aguilar is co-funded by the program FRANCE2030 QuanTEdu (Grant ANR-22-CMAS-0001) and CNES. 

\section*{Acknowledgments}
The authors would like to thank A. M. Akulshin, E. de Clercq, I. G. Hughes, M. Lepers and F. S. Ponciano-Ojeda for fruitful discussions. This work was partly supported by the french RENATECH network and its FEMTO-ST technological facility (MIMENTO).

\section*{Disclosures}
The authors declare no conflicts of interest.

\section*{Data availability statement}
The data that support the findings of this study are available from the corresponding author upon reasonable request.

\end{document}